\documentclass[sigconf]{acmart}

\AtBeginDocument{%
	}

\usepackage{booktabs} % For formal tables

\usepackage{amsmath}
\usepackage{balance}

\usepackage{graphicx}
\usepackage{array}

\usepackage{psfrag,subfigure,float}
\usepackage{algorithm,algorithmic}
\usepackage{diagbox} 
\usepackage{hhline}
\usepackage{caption}

\usepackage[most]{tcolorbox}

\makeatletter
\def\BState{\State\hskip-\ALG@thistlm}
\makeatother

\setcopyright{acmcopyright}
\copyrightyear{2024}
\acmYear{2024}
\setcopyright{acmlicensed}\acmConference[MM '24]{Proceedings of the 32st ACM International Conference on Multimedia}{October 28-November 1, 2024}{Melbourne, Australia}
\acmBooktitle{Proceedings of the 32st ACM International Conference on Multimedia (MM '24), October 28-November 1, 2024, Melbourne, Australia}
\acmPrice{15.00}
\acmDOI{10.1145/3581783.3612843}
\acmISBN{979-8-4007-0108-5/23/10}

\begin{document}
% \fancyhead{}

\title{Revisiting Vision-Language Features Adaptation and Inconsistency for Social Media Popularity Prediction}

\author{Chih-Chung Hsu}\thanks{This research was supported in part by Higher Education Sprout Project and High Education Humanities and Social Science Subjects Benchmark Project, Ministry of Education to the Headquarters of University Advancement at National Cheng Kung University (NCKU)
}
\affiliation{%
  \institution{Institute of Data Science, National Cheng Kung University}
      \country{Tainan, Taiwan}
  %\streetaddress{1, Daxue Rd., East Dist.}
  %\city{Tainan, Taiwan} 
  %\postcode{70101}
}
\email{cchsu@gs.ncku.edu.tw}

\author{Chia-Ming Lee}
\affiliation{%
	\institution{Institute of Data Science, National Cheng Kung University}
     \country{Tainan, Taiwan}
    %\streetaddress{1, Daxue Rd., East Dist.}
    %\city{Tainan, Taiwan} 
    %\postcode{70101}
}
\email{zuw408421476@gmail.com}

\author{Yu-Fan Lin}
\affiliation{%
	\institution{Institute of Data Science, National Cheng Kung University}
     \country{Tainan, Taiwan}
    %\streetaddress{1, Daxue Rd., East Dist.}
    %\city{Tainan, Taiwan} 
    %\postcode{70101}
}
\email{aas12as12as12tw@gmail.com}

\author{Yi-Shiuan Chou}
\affiliation{%
	\institution{Institute of Data Science, National Cheng Kung University}
     \country{Tainan, Taiwan}
    %\streetaddress{1, Daxue Rd., East Dist.}
    %\city{Tainan, Taiwan} 
    %\postcode{70101}
}
\email{nelly910421@gmail.com}

\author{Chih-Yu Jian}
\affiliation{%
	\institution{Institute of Data Science, National Cheng Kung University}
     \country{Tainan, Taiwan}
    %\streetaddress{1, Daxue Rd., East Dist.}
    %\city{Tainan, Taiwan} 
    %\postcode{70101}
}
\email{ru0354m3@gmail.com}

\author{Chi-Han Tsai}
\affiliation{%
	\institution{Institute of Data Science, National Cheng Kung University}
    %\streetaddress{1, Daxue Rd., East Dist.}
    \country{Tainan, Taiwan}
    %\postcode{70101}
}
\email{fateplsf567@gmail.com}

\renewcommand{\shortauthors}{Chih-Chung Hsu, Chia-Ming Lee, Yu-Fan Lin, Yi-Shiuan Chou, Chih-Yu Jian, \& Chi-Han Tsai}

\settopmatter{printacmref=false} 
\renewcommand\footnotetextcopyrightpermission[1]{}

\begin{abstract}

% Social-media popularity (SMP) prediction is a complex task as it involves user identities and the integration of various modalities data, such as post descriptions, images, geographical information, and temporal information. 
% %
% %To fully utilize the data, capturing the key latent factor and the interaction between different features is essential for this task. 
% %
% In recent years, many researchers have combined pre-trained vision-language (VLM) model (e.g. CLIP, ALBEF), which learned the amount of knowledge from large dataset, thereby archiving high-quality image and text representations for SMP task. 
% %
% Afterwards, VLM-adapters or feature alignment is employed to adapt these vision-language features to popularity, boosting model performance and yielding promising results.
% %
% However, user posts on social media platforms often exhibit insignificant correlations between content, such as differing semantic information between images and post descriptions. Even though adapting VLM features to the SMP task have been widely regarded as the effective approach for improving model performance, there is still a lack of comprehensive analysis. 
% %
% In this paper, we provide extensive analysis to illustrate the changes and effectiveness of VLM-features when adapted to different post qualities and popularity scores.

Social media popularity (SMP) prediction is a complex task involving multi-modal data integration. While pre-trained vision-language models (VLMs) like CLIP have been widely adopted for this task, their effectiveness in capturing the unique characteristics of social media content remains unexplored. This paper critically examines the applicability of CLIP-based features in SMP prediction, focusing on the overlooked phenomenon of semantic inconsistency between images and text in social media posts. Through extensive analysis, we demonstrate that this inconsistency increases with post popularity, challenging the conventional use of VLM features. We provide a comprehensive investigation of semantic inconsistency across different popularity intervals and analyze the impact of VLM feature adaptation on SMP tasks. Our experiments reveal that incorporating inconsistency measures and adapted text features significantly improves model performance, achieving an SRC of 0.729 and an MAE of 1.227. These findings not only enhance SMP prediction accuracy but also provide crucial insights for developing more targeted approaches in social media analysis.

\end{abstract}
\maketitle



\section{Introduction}

% With the development of the internet, communication among people increasingly relies on social-media platforms, which have become indispensable channels for understanding the world and interacting with others. Due to their importance and ubiquity, social media platforms have gradually become inseparable from commercial advertising, giving rise to fields such as recommendation systems and social network analysis. Accurately predicting the popularity of posts even enhance the effectiveness of algorithms such as recommendation systems \cite{recomment}, thereby improving the efficiency of commercial advertising applications and reducing development costs \cite{reocommend2}.

The ubiquity of social media platforms has revolutionized how we communicate, understand the world, and interact with others. These platforms have become integral to various domains, particularly in commercial advertising and marketing. Consequently, the ability to accurately predict the popularity of social media posts has emerged as a critical task, with significant implications for recommendation systems, content strategy, and advertising efficiency \cite{reocommend2, recomment}. However, the complexity of this task lies in the multi-modal nature of social media content, where the relationship between visual and textual elements is often overlooked. This paper critically examines the effectiveness of widely-used Vision-Language Models (VLMs), particularly CLIP, in capturing the unique characteristics of social media content for popularity prediction.

Social media platforms host millions of users, and the task of SMP prediction aims to forecast how many people will view a user's post. Compared to other multi-modal learning tasks, such as modality-wise feature translation or generation \cite{imagesearch1,imagesearch2}, and vision-question-answering (VQA) \cite{VQA}, social media prediction is relatively complex. Because a post typically includes various modalities of data, such as geographical information, temporal data, images, text, hashtags, and emojis. Wu \emph{et al.} introduced SMP dataset \cite{smp1,smp2,smp3,SMP2023}, which collected from Flicker and including the huge numbers of posts from different users around the world, to develop related researches and applications.

To address SMP prediction using features extracted from social media, traditional tree-based regression schemes have been employed, including XGBoost \cite{xgboost}, Random Forest \cite{rfr}, LightGBM \cite{lightgbm}, CatBoost  \cite{prokhorenkova2018catboost}, and even model stacking strategy \cite{stacking}. Nevertheless, these tree-based methods \cite{2022Catboost,hsu2017social,hsu2018social,hsu2019social,hsu2020social} have encountered a performance bottleneck in the SMP task because it relies heavily on feature extraction but cannot effectively integrate the semantic information hidden in different modalities, especially when encountering unseen data (a.k.a. predicting a post's popularity from new users). Besides, the importance of data splitting and external user-related features are also discussed in prior-works \cite{hsu2022,Hyfea,hsu2023}.

With the immense success of vision-language models (VLMs), deep learning has achieved significant breakthroughs. By collecting large amounts of image-text pairs and using contrastive learning \cite{SimCLR} methods to align or fuse features from different modalities \cite{chen2023double,2022Chen}, these models provide high-quality and semantically rich features, effectively enhancing the performance and generalizability of downstream models. Previous researches \cite{2022Chen,2022Wu,2023sota,hsu2023,chen2023double} has shown that adapt a knowledge from VLMs, such as CLIP \cite{CLIP}, to obtain features between images and text descriptions from user posts can effectively improve the performance of SMP prediction models.

Recently, Chen \emph{et al.} introduced the TTC-VLT \cite{2022Chen}, combining two pre-trained vision and language transformers and integrating title-tag contrastive learning for title-visual and tag-visual, which separately extracts multimodal information from two types of text. Subsequently,  DFT-MOVLT \cite{chen2023double} was proposed. With the design of multi-objective pre-training, significantly enhancing the performance of social media popularity prediction. Zhang \emph{et al.} focused more on the dependencies between posts from the same user. They proposed the Dependency-aware Sequence Network (DSN) \cite{2023sota}, which fully leverages the importance of post content in social media popularity prediction. This framework combines a CLIP-adapter \cite{clipadapter} by adding a small number of learnable layers while retaining the original backbone of CLIP, enhancing the representation capability for specific tasks during fine-tuning.

However, despite their effectiveness, whether these models accurately capture the corresponding relationships remains an open question. According to our observations, there are often large variations in posting habits between different user groups, and these characteristics have lacked attention in previous studies. In the SMP dataset, those posts with very low popularity (popularity score is lower than 2) often come from new users, and most new users do not pay attention to the quality of text descriptions (e.g., text descriptions consisting of meaningless characters or hard-to-encode emojis). Posters of very popular posts (popularity score is larger than 10) are more likely to focus on specific topics, or are more likely to have accounts carefully created for marketing.

The posts between the two are the majority of the posts that make up the SMP dataset. Among them, we found that the trend of image-text similarity will decrease with the increase in post-popularity, which means that posts with inconsistent images and texts may increase with popularity. 

Finally, when reviewing the purpose of feature alignment \cite{albef,chen2019progressive} or knowledge adaptation \cite{KnowledgeAdaptation}, we aim for the latent semantics of features extracted from VLMs to be aligned with or adapted to popularity. Although these strategies are widely regarded as effective in SMP tasks, there remain numerous issues that need to be explored. In this paper, we address the aforementioned questions by conducting experiments to compare the relevance of VLM-features in the feature space. We specifically analyze whether the user is a celebrity and the relative quality of the images and text in the posts. The main contribution of this paper are three-folds:
\begin{itemize}
	\item We observed that the phenomenon of mismatched text and images is a widely overlooked issue in the SMP task. We examined the text and image features of the SMP dataset. The semantic inconsistency between image and text increases along with popularity scores.
	\item We consult the comprehensive analysis for semantic inconsistency, in term of textual feature quality, across different popularity intervals and the impact of VLM features adaptation on the SMP task.
    \item Based on analysis results, we provide the further experiments and design a helpful features for enhancing model performance.
\end{itemize}
The rest of this paper is organized as follows. Section 2 illustrates the semantic inconsistency hypothesis of image-text features for the SMP dataset. Section 3 and 4 present our analysis and experiment results, respectively. Finally, the conclusion is drawn in Section 5.

\begin{figure*}
	\centering
	\includegraphics[width=1\textwidth]{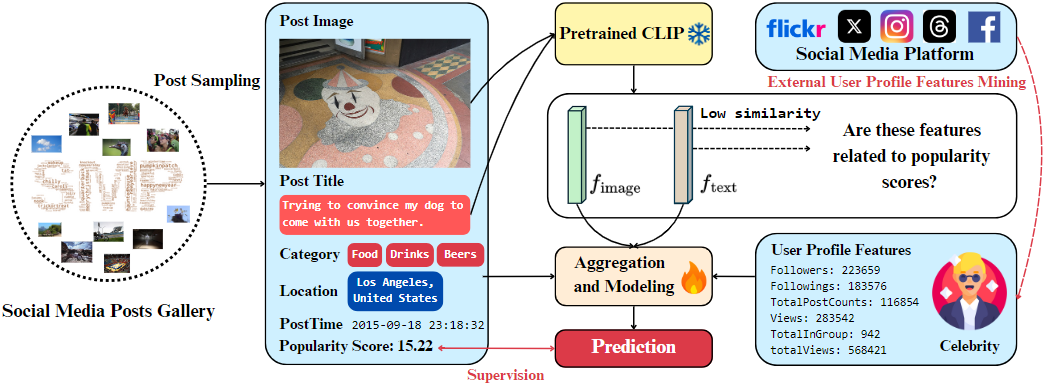}
	\caption{The architecture commonly used for social media popularity prediction models is illustrated below. For every posts, Image and text descriptions are first encoded by the pre-trained Vision-Language Models (VLMs) like CLIP to extract features. These features are then aggregated with other post features (such as spatiotemporal features) and user-related features to serve as the model's input. Despite the effectiveness for SMP task, the vision-language features adaptation need to be explored, especially with semantic inconsistency between image and text description from social post.}
	\label{fig:SMPD24.png}	
\end{figure*}

%\hypbox{The Multitask Scaling Hypothesis}{%
%There are fewer representations that are competent for $N$ tasks than there are for $M<N$ tasks. As we train more general models that solve more tasks at once, we should expect fewer possible solutions.
%}

%%%%%%%%%%%%%%%%%%

\section{Semantic Inconsistency Hypothesis}

In this section, we define the observed phenomena and formally propose hypotheses, focusing on the phenomenon of mismatched text and images and its potential impacts.

%\subsection{Inconsistency Between Image and Text from Social Posts}
\textbf{Background. }Previous studies have highlighted that using VLMs to extract features from images or post description (such as 'Title' and 'Tags') \cite{2022Chen} from social-media posts, and aligning these features or using additional adapters to adapt the features to the target domain's distribution \cite{2023sota,chen2023double,2022Wu}, is an effective solution for enhancing the SMP prediction.

However, from the another perspective, there is a common phenomenon of image-text inconsistency on social-media platforms. \textbf{Users might employ post descriptions that are completely unrelated to the images they posted}, as illustrated in Figure \ref{fig:SMPD24.png}. Additionally, users' activity degree may influence their preferences for the post descriptions they use. Therefore, in this paper, we conduct a comprehensive investigation of the SMP dataset to understand the prevalence of image-text inconsistency within the dataset.

\textbf{Potential Impact and Limitation. } The phenomenon of image-text inconsistency has been largely overlooked in previous researches. 

%Using CLIP-adapter \cite{clipadapter} or contrastive learning \cite{albef} to enhance model performance is primarily based on projecting the originally encoded feature representations into relatively low-dimensional sub-spaces and fine-tuning these features only within those sub-spaces. This approach aims to avoid retraining from scratch or catastrophic forgetting, thereby reducing the difficulty of effectively aligning the feature representations of images or texts with popularity predictions.

If image-text inconsistency exists, several questions may need to be discussed. For instance, despite the success of SMP task in previous works \cite{2022Chen,2023sota,chen2023double}, these models might over-fit to the training set, posing the lack of generalizability when the dataset changes or real-world deployment.

In additional, the inconsistency between image and post descriptions may constraint the model performance and data-efficiency. Especially with low-quality image or post description. To elaborate, VLM-features may fail to align the semantic information with a popularity score of post during feature adaptation.

%%%%%%%%%%%%%%%%%%%%%%%%%%%%%%%%%%
\section{Analysis Against The Hypothesis}

In this section, we aim to validate the hypotheses proposed in the previous section through three perspectives. 

\begin{itemize}
	\item First, we will observe the trend of  similarity among posts within different popularity intervals to explore the overall trend of inconsistency between image and text from a social posts.
	\item Next, we visualize the distribution of perplexity to examine the quality trends of posts across different popularity intervals. 
	\item Finally, we analyze the changes before and after VLM representations adaptation to figure out the difference of image-text features during adaption for aligning popularity score.
\end{itemize}
Additionally, we aim to minimize the importance of external user information in our analysis, even though it is widely regarded as important in SMP prediction tasks \cite{Hyfea,hsu2023}.

\begin{figure}
	\centering
	\includegraphics[width=0.45\textwidth]{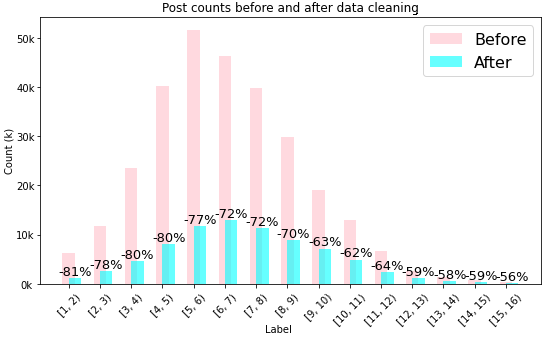}
	\caption{The number of training data used for analysis after data cleaning. The proportion of removed posts decreases as popularity increases, implying that the proportion of low-quality posts is relatively higher among less popular posts.}
	\label{fig:dataclean.png}	
\end{figure}

\begin{figure}
	\centering
	\includegraphics[width=0.45\textwidth]{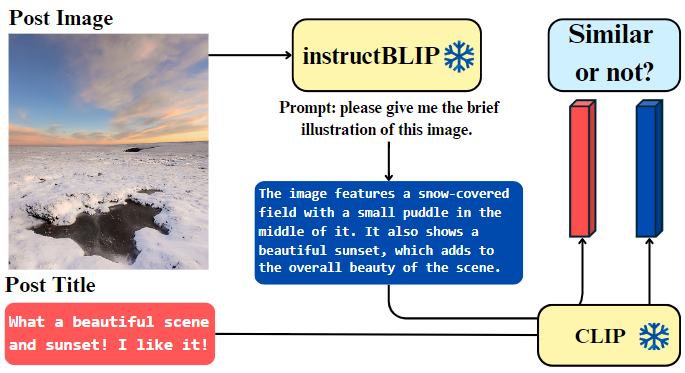}
	\caption{The pipeline of calculating similarity between image and post description ('Title') from a post. We use a pre-trained VLM model to obtain image caption and calculate their similarity to post descriptions. This similarity serves to evaluate the semantic consistency between image and post descriptions.}
	\label{fig:similarity_pipeline.png}	
\end{figure}

\begin{figure}
    \centering
    \subfigure[Similarity between Title and Caption.]{
        \label{adapted_tsne.tif}
        \includegraphics[width=0.228\textwidth]{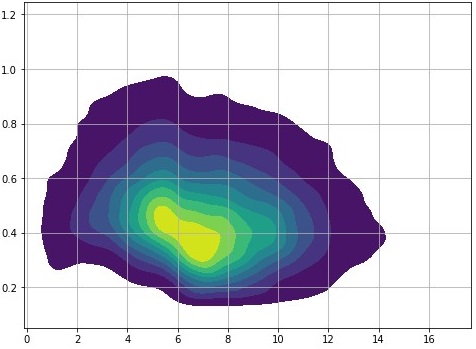}}
    \subfigure[Perplexity Score of Title.]{
        \label{adapted_tsne}
        \includegraphics[width=0.235\textwidth]{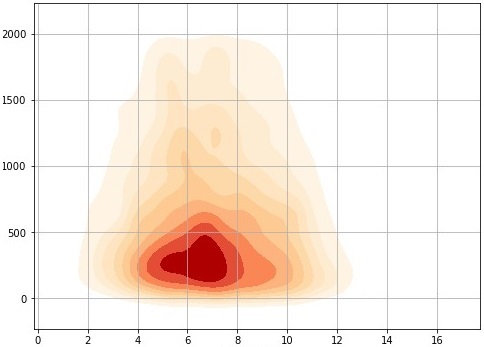}}
    \caption{Cosine similarity between title and image caption features is shown on the left. The calculation process is illustrated in Figure \ref{fig:similarity_pipeline.png}. %This distribution helps in understanding the alignment or misalignment between textual and visual content in social media posts. 
    The right image shows the perplexity score of post descriptions, measured by pre-trained GPT-2. %Perplexity is a metric used to evaluate the fluency and coherence of text, with lower values indicating higher quality. This analysis provides insights into how well the titles are constructed and their potential impact on model performance.
    }
    \label{Fig.main}
\end{figure}

\begin{figure}
    \centering
    \subfigure[t-SNE of original CLIP-features.]{
        \label{adapted_tsne.tif}
        \includegraphics[width=0.23\textwidth]{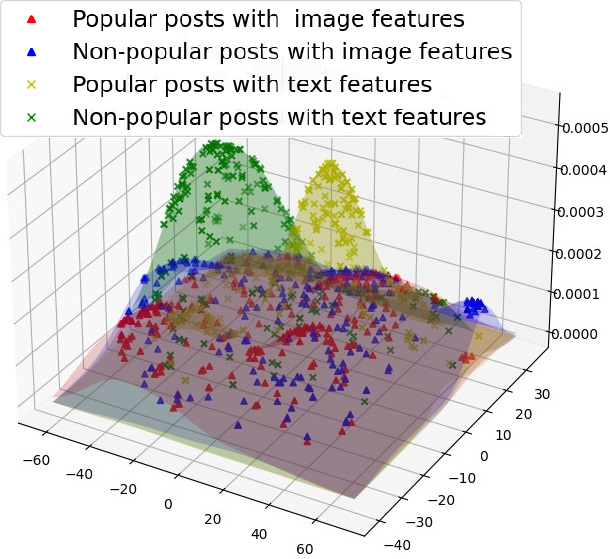}}
    \subfigure[t-SNE of adapted CLIP-features.]{
        \label{adapted_tsne}
        \includegraphics[width=0.23\textwidth]{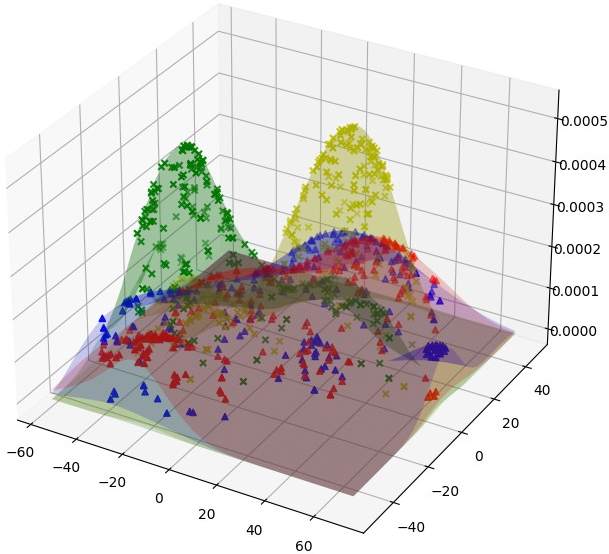}}
    \caption{3D t-SNE plot visualization of image-text features extracted by CLIP before and after adaptation, color means the density of post. Vision-language representations adaptation has a more significant impact on text features.}
    \label{Fig.tsne}
\end{figure}

\subsection{Low-quality Posts Removal for Analysis}

To conduct an effective analysis, we first performed a simple data preprcessing on the raw training data to ensure the quality of the posts used for analysis. Initially, we removed posts containing titles with excessive punctuation or emojis. Next, we eliminated posts with titles shorter than 20 characters. Finally, we excluded posts with title perplexity greater than 2000, calculated using a pre-trained GPT-2 model \cite{gpt2}. The remaining posts, therefore, possess relatively rich and meaningful semantic information, thereby enhancing the effectiveness of the analysis. The result is shown in Figure \ref{fig:dataclean.png}.

\subsection{Image-Text Inconsistency}

% To measure image-text inconsistency, we use InstructBLIP \cite{instructblip} to generate image descriptions (prompt: "please give me a short description of this image"). These image descriptions, along with the 'Title', are then used to extract features using a pre-trained CLIP \cite{CLIP}, as shown in Figure \ref{fig:similarity_pipeline.png}. The cosine similarity between the image description features and the text features is calculated to assess the degree of inconsistency between the image and post description.

To quantify the semantic inconsistency between images and text in social media posts, we develop a novel two-step approach. First, we employ InstructBLIP \cite{instructblip} to generate objective descriptions of the images, using the prompt "please give me a short description of this image". This step aims to capture the core visual content without bias from the associated text. Subsequently, we leverage a pre-trained CLIP model \cite{CLIP} to extract high-dimensional feature vectors for both the generated image descriptions and the original post titles. The degree of semantic inconsistency is then measured by calculating the cosine similarity between these feature vectors, as illustrated in Figure 3. This method allows us to systematically assess the alignment between visual and textual content across a large corpus of social media posts, providing insights into how this inconsistency varies with post popularity and potentially impacts the effectiveness of VLM-based features in SMP tasks.

As Figure \ref{adapted_tsne.tif} shown, we use a 3D kernel density plot to visualize the image-text similarity of all filtered posts in the SMP dataset. It can be observed that most posts are concentrated in the popularity range of 3 to 9, which we will refer to as the \textbf{main-trends posts interval} (as it better represents the overall trend). The other two ranges can be referred to as the \textbf{non-popular posts interval} and the \textbf{popular posts interval}. In the main-trends interval, we can observe a trend where the text-image similarity decreases as popularity increases, suggesting that these posts, which contain mismatched image and post description, are more likely to become popular within this interval. Since the samples in the popular and non-popular posts interval are limited, we need to carefully analyze them based on the text quality of their posts. We will discuss it in the next subsection.

%%%%%%%%%

\subsection{Textual Quality of Post Description}

%%%%%%
In the main-trends posts interval, the majority of posts show a decrease in perplexity as popularity increases, which is consistent with our observations in the section 3-2. These accounts are more likely to be active users, where the posts may have mismatched image and text semantically.

On the other hands, in the non-popular posts interval, the quality of the text descriptions is relatively low, as most low-quality posts were removed as section 3-1. As for the popular posts interval, we can infer from Figures \ref{fig:dataclean.png} and \ref{adapted_tsne} that posts in this interval are more likely from accounts used for specific purposes rather than by individual users, such as fan-pages. This is indicated by the relatively lower number of posts, longer and higher-quality text descriptions, and the relative rarity of mismatched image and text compared to the main-trends interval.

\subsection{Effectiveness of Vision-Language Representations Adaptation}

\hspace{\parindent} \textbf{Experiment Setting.} We use CLIP-adapter \cite{clipadapter} to transfer the information from pre-trained CLIP to the SMP task, aiming to align the image-text representation as closely as possible with the distribution of popularity semantically. Specifically, we employ two independent CLIP-adapters for the image features and text information ('Title'), respectively, setting the residual coefficients to 0.2 and 0.6. These settings are consistent with \cite{2023sota}.

We used t-SNE \cite{vandermaaten08a} to visualize the image and text features of all posts in the non-popular posts interval and popular posts interval, as shown in the figure. After adaptation, the image features did not exhibit significant changes. This can be attributed to the susceptibility of images to noise, quality, semantics, and other factors, suggesting that image features may not be critical for the SMP task. In contrast, the text features showed relatively significant changes. The text features of low-quality posts and high-quality posts became more convergent and aligned more effectively with popularity levels.

\subsection{Explanation for the Analysis}

The phenomenon of mismatched text and images might be a widely overlooked issue in the SMP task. In this paper, we conducted extensive experiments and analyses on text and image information to understand the occurrence of mismatched text and images, the distribution of text description quality, and the impact of VLM feature adaptation on the SMP task.

The mismatched text and image phenomenon and the quality of text descriptions can roughly describe user preference trends across different popularity intervals. Analyzing the behavior of different groups may provide slight improvements in performance. From a feature analysis perspective, text features may play a relatively important role when combining unstructured data such as text and images, as their semantics or quality can more easily align with varying degrees of popularity. In contrast, image features are more challenging to align due to their susceptibility to image quality and their more diverse distribution in the feature space, making it difficult to align them with popularity levels.

%%%%%%%%%%%%%%%%%%%%%%%%%%%%%%%%%%
\section{Experiment}

In this section, we aim to extend analysis results from previous section to improve SMP task. We stack perplexity and similarity scores as features, allowing the model to utilize the inconsistency degree of images and text, as well as the impact of text quality on popularity. We also replaced the VLM-features used in \cite{hsu2023} with those adapted to align with popularity score.

\subsection{Dataset}

We use the Social Media Prediction Dataset (SMPD), derived from \cite{smp1}, including 486,194 posts from 60,093 users. As per \cite{hsu2020social}, the treatment of missing data is assigned. We keep the default setting at \cite{SMP2023} and divide the training set and test set according to the timestamp of posts, and their numbers are 305,613 and 180,581 respectively.

\subsection{Evaluation Metrics}To assess prediction performance, we utilize the precision metric Mean Absolute Error (MAE) and the correlation metric Spearman Ranking Correlation (SRC), as described in \cite{rc}. For a dataset with k samples, where the ground-truth popularity set is denoted by $S$ and predicted popularity set $\hat{S}$ both ranging from 0 to 1, the MAE is calculated as follows:
\begin{equation}
M A E=\frac{1}{k} \sum_{i=1}^n\left|\hat{S}_i-S_i\right|
\end{equation}
The SRC metric is employed to determine the ranking correlation between 
 $\hat{S}$ and $S$:
\begin{equation}
S R C=\frac{1}{k-1} \sum_{i=1}^k\left(\frac{S_i-\bar{S}}{\sigma_S}\right)\left(\frac{\hat{S}_i-\overline{\hat{S}}}{\sigma_{\hat{S}}}\right)
\end{equation}
A lower MAE and a higher SRC indicate better performance.

\begin{table}
    \centering
        \caption{The user information (U), category variable (C), image-textual features (I \& T), generated features used for prediction (P \& S). The data entries with $*$ are generated by the methods mentioned as Section 3.}
	\scalebox{0.95}{
    \begin{tabular}{lrr}
        \toprule
        Data Entry &  Description\\
        \midrule
        Uid & The user this post belongs to.\\
        Ispublic &  Is the post authenticated with 'read' permissions.\\
        Ispro & Is the user belong to pro member.\\      
        Latitude & The latitude of the posting location.\\ 
        Longitude & The longitude of the posting location.\\
        GeoAccuracy & The accuracy level of the location information.\\
        Postdate & The publish timestamp of the post.\\
        \midrule
        Category & The main category on the posting \\
        Subcategory & The sub-category on the posting \\
        Concept & The concept on the posting \\
        \midrule
        Image & The image a user posted \\
        Title & The text description of the posting \\
        Tag & The hashtags of the posting \\
        \midrule
        $\rm{Perplexity^*}$ & The number of people the user follows.\\
        $\rm{Similarity^*}$ & The number of followers of the user.\\
        \bottomrule
    \end{tabular}
    }
    \label{tab:metadata}
\end{table}

\subsection{Base Model and Feature Engineering}

\hspace{\parindent} \textbf{Base-model.} To keep experiment simple and easy to analysis, we use the GBTN \cite{hsu2023} for our experiments. This is the ensemble model which takes weighted average of LightGBM \cite{lightgbm} and TabNet \cite{tabnet}. The hyper-parameter settings and data-splitting strategy are the same as \cite{hsu2023} illustrated.

\textbf{Feature Engineering.} The features we used are listed as Table-\ref{tab:metadata}. We use the pre-trained CLIP \cite{CLIP} to extract image-text ('Title' and 'Tags') features. Hierarchical Category Embedding (HCE) \cite{2023sota} is used to capture hierarchical relationship among category variables. We use CLIP-adapter \cite{clipadapter} to align image/text features with popularity scores in offline. HyFea (H) \cite{Hyfea} features are also used in the experiment. The final inputs can be defined as $f_{all}$ = [$f_{U}$, $f_{C}$, $f_{H}$, $f_{I'}$, $f_{T'}$, $f_{P}$, $f_{S}$], where $[\cdot]$ means stacking operation. The predicted popularity score $\hat{S}$ can be written as:

\begin{equation}
\hat{S} = \alpha \cdot \text{LigheGBM}({f_{all}}) + (1-\alpha) \cdot \text{TabNet}({f_{all}}),
\end{equation}
where $\alpha$ is set to 0.7, so as to aggregate the different information which learned by these two type of models.

\begin{table}[htbp]
	\centering
	\caption{Performance comparison of base model with stacking different features.}
	\label{tab:proposed method}
	\scalebox{0.78}{
		\begin{tabular}{ccc|c|c|ccc}
			\hline
			& & \multicolumn{2}{c}{LGBM} & \multicolumn{2}{c}{TabNet} \\
			& & SRC $\uparrow$ & MAE  $\downarrow$& SRC $\uparrow$ & MAE$\downarrow$ \\
			\hline
			\multicolumn{1}{l}{Baseline (U+C)} & & 0.636 & 1.436 & 0.625 & 1.502 \\
            \hline
            \multicolumn{1}{l}{\textbf{+ External User Feature (H)}} & & 0.707 & 1.261 & 0.66 & 1.362 \\
            \multicolumn{1}{l}{\textbf{+ Hierachical Category Embedding (HCE)}} & & 0.712 & 1.257 & 0.678 & 1.342 \\
			\multicolumn{1}{l}{Image and Text Feature (I+T)} & & 0.71 & 1.254 & 0.679 & 1.344 \\
			\multicolumn{1}{l}{Adapted Image Feature (I')} & & 0.711 & 1.255 & 0.676 & 1.346 \\
			\multicolumn{1}{l}{Adapted Text Feature (T')} & & 0.715 & 1.248 & 0.683 & 1.334 \\
            \multicolumn{1}{l}{\textbf{+ Adapted Image and Text Feature (I'+T')}} & & 0.718 & 1.249 & 0.692 & 1.331 \\
            \multicolumn{1}{l}{Perplexity Score (P)} & & 0.722 & 1.238 & 0.698 & 1.316 \\
            \multicolumn{1}{l}{Similarity Score (S)} & & 0.719 & 1.244 & 0.694 & 1.328 \\
            \multicolumn{1}{l}{\textbf{+ Perplexity and Similarity Score (P+S)}} & & 0.724 & 1.231 & 0.704 & 1.298 \\
			\hline
		\end{tabular}
	}
\end{table}

\subsection{Results and Ablation Study}

Experiment results indicate that adding perplexity and similarity features can improve model performance. Furthermore, the results show that the adapted text features have a greater impact than image features. This implies that after adaptation, the semantic content of text features aligns better with the distribution of popularity, which is consistent with our analysis in the previous chapter. Additionally, similarity features also contribute to performance improvement, especially when combined with perplexity. After model ensemble, we achieved final results with an SRC of \textbf{0.729} and an MAE \textbf{1.227}, which is better than the previous results \cite{hsu2023}.

\section{Conclusion}

In this paper, we first identified the semantic inconsistency between image and post description in the SMP dataset, noting that this inconsistency might become more frequent with increasing popularity. We demonstrated this through analysis and visualization, prompting a further investigation into the widely used VLM features (notably CLIP) and their adaptation. 

In these analysis, we figure out some of factors are important and can be utilized for SMP task. For instance, the textual quality of post descriptions across different popularity interval, and the similarity score between image and post description, which are overlook in previous works. Finally, our experiments demonstrated that incorporating these feature significantly improves the model performance.

\bibliographystyle{ACM-Reference-Format}
\balance
\newpage
\bibliography{smpd23_bibliography} 

\end{document}